\documentclass[epj]{svjour}
\usepackage{epsf}
\begin{document}
\title{
Numerical analysis
of the dissipative two-state system
with the density-matrix Hilbert-space-reduction algorithm
}
\titlerunning{Numerical analysis of the dissipative
two-state system}
\author{Yoshihiro Nishiyama}
\institute{
Department of Physics, Faculty of Science,
Okayama University,
Okayama 700-8530, Japan}
\date{Received: date / Revised version: date}
%
\abstract{
Ground state of the dissipative two-state system
is investigated by means of the Lanczos diagonalization method.
We adopted the Hilbert-space-reduction scheme proposed by
Zhang, Jeckelmann and White so as to 
reduce the overwhelming reservoir Hilbert space 
to being tractable in computers.
Both
the implementation of the algorithm
and the precision applied for the present system
are reported in detail.
We evaluate the
dynamical susceptibility (resolvent) with
the continued-fraction-expansion formula.
Through analysing the resolvent over a frequency range,
whose range is often called `interesting' frequency, 
we obtain the damping rate and the oscillation
frequency.
Our results agree with those of a
recent quantum Monte-Carlo study, which
concludes that the critical dissipation from oscillatory
to over-damped behavior decreases as the
tunneling amplitude is strengthened.
\PACS{
{75.40.Mg} 
{numerical simulation studies} \and
{05.40.+j}
{Fluctuation phenomena, random processes, and 
Brownian motion} \and
{05.70.Ln}
{Nonequilibrium thermodynamics, irreversible
processes}
     } 
} 
\maketitle
\section{Introduction}
\label{section1}
Effect of dissipation on quantum tunneling
phenomenon
lies out of the scope of the conventional
weak coupling (Markovian) approximation,
and has been studied extensively so far
\cite{Caldeira83b,Leggett87}.
In order to introduce dissipation,
Caldeira and Leggett \cite{Caldeira81,Caldeira83a}
involved
a thermal reservoir which consists of
oscillators influencing stochastic (Brownian) fluctuations
to the tunneling two-level system.
Their model, the so-called spin-boson model, is given by the 
following Hamiltonian,
\begin{equation}
{\cal H}=-\frac{\Delta}{2}\sigma^x+\sum_{i=1}^{N}
\omega_i a^\dagger_i
a_i +\frac{\sigma^z}{2}
\sum_{i=1}^{N}
f_i (a^\dagger_i+a_i),
\label{Hamiltonian}
\end{equation}
where the operators $\{\sigma^\alpha\}$ denote the Pauli operators 
and 
the operators
$a_i$ and $a_i^\dagger$ denote the 
bosonic annihilation and
creation operators, respectively, for the $i$-th oscillation
mode
($i=1 \sim N$).
The set of these oscillators works as the above-mentioned reservoir
with respect to the spin $1/2\mbox{\boldmath $\sigma$}$.
The coupling coefficients $\{ f_i \}$
should be arranged so as to satisfy
the so-called Ohmic-dissipation condition,
\begin{eqnarray}
J(\omega)&=&\pi\sum_i f^2_i \delta(\omega-\omega_i) \\
\label{ohmic_condition}
         &=&
\left\{
\begin{array}{lr}
2\pi\alpha\omega & (\omega\ll\omega_{\rm c}) \\
0 & (\omega\gg\omega_{\rm c})
\end{array}
\right. .
\end{eqnarray}
The dimensionless constant $\alpha$ controls
the strength of the dissipation, and
the cut-off frequency $\omega_{\rm c}$ defines the 
energy unit throughout this paper; $\omega_{\rm c}=1$.
The transverse field $\Delta$ induces quantum coherence
(tunneling amplitude) between the spin up-and-down states,
whereas the coupling to the reservoir 
disturbs the coherence.
These conflicting effects are the central concern of the
problem, which are apparently non-perturbative in nature.

\begin{figure}[htbp]
\begin{center}\leavevmode
\epsfxsize=8.5cm
\epsfbox{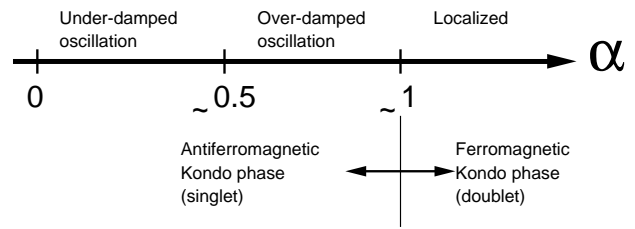}
\end{center}
\caption{
The ground-state phase diagram of the spin-boson
model ({\protect \ref{Hamiltonian}})
with the Ohmic dissipation.
Quantitative
estimation of
the relaxation parameters such as the
damping rate and the oscillation frequency is
the main concern of this paper.
}
\label{phase_diagram}
\end{figure}

It is noteworthy that
the 
spin-boson model (\ref{Hamiltonian})
is 
equivalent to the anisotropic 
Kondo model \cite{Chakravarty82,Bray82,Hakim84,Guinea85}.
The parameter $\alpha$ controls the strength of the
longitudinal Kondo coupling, whereas $\Delta$ controls the
transverse-coupling strength.
Thereby,
the region $\alpha<1$ ($\alpha>1$) is identified
as the the antiferromagnetic 
(ferromagnetic) Kondo phase; see the ground-state phase diagram shown
in Fig. \ref{phase_diagram}.
That is, in the region $\alpha<1$, the up-and-down spin states
form a coherent (singlet) state through a certain tunneling amplitude,
whereas in $\alpha>1$, the tunneling amplitude vanishes
owing to the dissipation \cite{DeRaedt83,DeRaedt84}.
Hence, in the region $\alpha>1$, the ground state is
degenerated doubly.
The effective tunneling amplitude is found to
vanish at the phase boundary in the form
\cite{Guinea85c,Guinea85b},
\begin{equation}
\label{effective_tunneling}
\Delta_{\rm eff}=\Delta
\left(
\frac{\Delta}{\omega_{\rm c}}
\right)^{\frac{\alpha}{1-\alpha}} .
\end{equation}
Owing to the equivalence, we readily investigate the 
equilibrium (thermodynamic) properties
by means of
various theoretical techniques which have been 
developed so far for the Kondo problem.
Note, however, the dynamical (non-equilibrium) properties
are rather out of the scope of these analytical techniques;
refer to the paper \cite{Lesage96}
for a recent 
analytical approach.
The dynamical properties, especially in the frequency
range $\sim \Delta_{\rm eff}$, are the very concern
of the present topic.
The noninteracting-blip approximation \cite{Leggett87,Chakravarty84}
was invented so as to
describe the relaxation dynamics of this particular frequency range.
This approximation is, however, justified in 
a rather limited parameter range $\alpha<0.5$.

Hence,
in order to investigate the relaxation properties,
various numerical simulations have been performed
so far.
Chakravarty and Rudnick performed quantum Monte-Carlo
simulation \cite{Chakravarty95};
the model they simulated is a one-dimensional
long-range Ising model,
which was derived
\cite{Anderson70,Anderson71}
through eliminating the
reservoir (conduction election) degrees of freedom.
They succeeded in obtaining the spectral function 
through the Pad\'e approximation of the imaginary time correlation function
followed by the analytic continuation (Wick rotation).
As a result, they found that the long-range asymptotic form 
of the dynamical correlation is governed by power law.
This conclusion was supported by Strong \cite{Strong97}
with use of the similar numerical technique.
V\"olker \cite{Voelker98}
followed the Chakravarty-Rudnick analysis, and 
reported that 
`quasi-particle'
picture
explains the spectral-function data.
As a consequence,
he obtained the damping rate
and frequency.
We utilize the picture to
analyze our
density-matrix-renormalization data.
Costi and Kieffer \cite{Costi96,Costi97}
used the Wilson numerical renormalization technique 
to calculate the spectral function.
Their technique is particularly useful 
in order to investigate
the Kondo fixed-point (very low temperature) physics
definitely.
Time-evolution simulation was performed with use
of stochastic sampling (path integral) algorithm
\cite{Stockburger98}.
The method has an advantage over others
that one can observe real-time
dynamics directly.
The stochastic-sampling error, however,
grows as the evolution time is lengthened.

In the present paper, for the first time,
we perform Lanczos-diagonalization
analysis of the spin-boson model (\ref{Hamiltonian}).
In order to command vast assembly of the
reservoir-oscillator modes, we adopted the density-matrix
truncation scheme proposed by Zhang, Jeckelmann and White
\cite{Zhang98}.
(They studied the one-dimensional polaron system with this
truncation technique.)
The rest of this paper is organized as follows.
In the next section, we propose
an implementation of their
algorithm to the spin-boson model (\ref{Hamiltonian}),
and report the precision in detail.
In Section \ref{section3},
by means of this new technique, we investigate
the relaxation properties of the spin-boson model.
We evaluate
the dynamical susceptibility (resolvent), which is
readily calculated in our scheme with use of
the continued-fraction-expansion formula \cite{Gagliano87,Gagliano88}.
Analyzing the analyticity of the resolvent,
we confirm the above-mentioned `quasi particle'
picture \cite{Voelker98} so as to
obtain the damping rate and frequency.
Our results are
contrasted with the former quantum Monte-Carlo
results \cite{Voelker98}.
In the last section, we give summary and discussions.

\section{
Application of the
Hilbert-space-truncation algorithm 
to the spin-boson model
}
\label{section2}

In this section, we propose a prescription for
adopting the Hilbert-space-reduction method to the
spin-boson model.
We then demonstrate the precision of our new scheme.

\subsection{Hilbert-space-reduction algorithm}
\label{section2_1}

Even a single oscillator (boson)
spans infinite-dimensional Hilbert space.
Therefore, in order to treat an oscillator
with the exact diagonalization method,
one must truncate the 
Hilbert space inevitably.
In conventional simulations, the
boson state is represented in terms
of its occupation number, and
those states whose occupation number exceeds
a limit are disregarded (discarded).
Zhang, Jeckelmann and White \cite{Zhang98}
proposed an alternative
representation and a truncation criterion.
Applying their scheme to a polaron system 
(one-dimensional Holstein model),
they demonstrated that the scheme yields
very precise results, although the dimensionality
of each oscillator is reduced to three.
This truncation is called `(numerical) renormalization.'
Their new bases are particularly efficient in those cases
where the oscillator equilibrium position is
shifted by a certain external force (coordinate-coordinate coupling).
This is precisely the case of the 
present model (\ref{Hamiltonian}).

\begin{figure}[htbp]
\begin{center}\leavevmode
\epsfxsize=8.5cm
\epsfbox{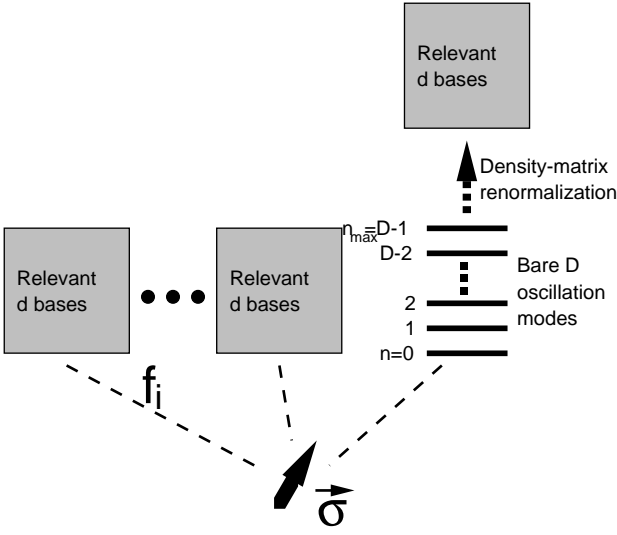}
\end{center}
\caption{
Schematic drawing of the
density-matrix-renormalization algorithm applied
to the spin-boson model ({\protect \ref{Hamiltonian}}).
}
\label{DMRG}
\end{figure}

In the following,
we explain the detail how we adopt their
algorithm to the spin-boson model.
First, we limit the 
Hilbert-space dimensionality of each oscillator to
$d$ dimensions except an oscillator with $D$ dimensions;
see Fig. \ref{DMRG}.
We call the $d$-dimensional oscillators `small oscillators,'
and the $D$-dimensional one `big oscillator.'
The choice of the big oscillator 
is to be made in sequence among the
various reservoir oscillator modes ($\alpha=1 \sim N$).
The sequence is continued until
the relevant bases, explained below, become converged.
In our experience, a few sweeps are sufficient for the convergence.
The space of each small oscillator
is spanned by the above-mentioned truncated
relevant bases, and thus the creation and the annihilation
 operators 
should be
represented in terms of these bases correspondingly.
(Before renormalization,
the space is allowed to be of the number representation
($|n\rangle$, $n=1 \sim d-1$), and the operators
are expressed in terms of this subspace.)
On the other hand,
the space of the big oscillator is spanned
by the occupation-number representation
with $n=0 \sim D-1$.
From this $D$-dimensional space, $d$ relevant bases are
renormalized (extracted)
in the following manner.

Second, we diagonalize the Hamiltonian (\ref{Hamiltonian})
with the Lanczos technique
to obtain the ground state $|\Psi_0 \rangle$.
Note that the Hamiltonian consists of the operators
whose matrix elements are already prepared in the above.
The ground-state vector should be represented
in the form,
\begin{equation}
| \Psi_0 \rangle
=
\sum_{i,j}
\psi_{ij}
|i\rangle_A |j\rangle_B,
\end{equation} 
where the bases $\{ |i\rangle_A \}$ are of the big site,
and the bases $\{ |j\rangle_B \}$ are of the remaining part of the
system.
Thereby, we construct the (reduced)
density matrix subjected to the big oscillator,
\begin{equation}
\rho_{i i'}=\sum_j \psi_{i j}\psi^{*}_{i' j}.
\end{equation}
The new bases $\{ {\bf u}^\beta \}$ are determined
so as to diagonalize the density matrix,
\begin{equation}
\rho {\bf u}^\beta = w_\beta {\bf u}^\beta.
\end{equation}
According to the proposal
\cite{White92,White93}, the new relevant bases (subspace)
should be spanned by the eigenvectors
 ${\bf u}^\beta$ up to the $d$-th
largest weight (eigenvalue) $w_\beta$.
That is, the new annihilation operator of the big oscillator
is
re-represented in terms of these $d$ relevant eigenvectors
in the following manner,
\begin{equation}
[\tilde{b}_i]_{\beta \beta'}={}^{{\rm t}}
{\bf u}^\beta b_i {\bf u}^{\beta'}  .
\end{equation}
Because the renormalization is subjected to the
reduced density matrix,
the renormalization is called `density matrix'
renormalization \cite{White92,White93}.

The renormalization is continued for every oscillator modes ($i=
1\sim N$)
successively until the relevant bases become fixed
(converged).
In our experience, a few sweeps are sufficient
for the convergence.

Finally, we mention how we chose the frequencies 
$\{ \omega_\alpha \}$ 
and the
coupling constants $\{ f_i \}$.
The chose is arbitrary under the constraint eq. 
(\ref{ohmic_condition}).
We have uniformly distributed the oscillator frequency
over the range $0<\omega<\omega_{\rm c} (=1)$,
and determined the coupling constants so as to satisfy the
constraint.
We numbered the oscillator modes ($\alpha=1 \sim N$)
in order of the frequency $\omega_\alpha$
(from the slowest to the fastest).

\subsection{Precision of the Hilbert-space-truncation algorithm}

\begin{figure}[htbp]
\begin{center}\leavevmode
\epsfxsize=8.5cm
\epsfbox{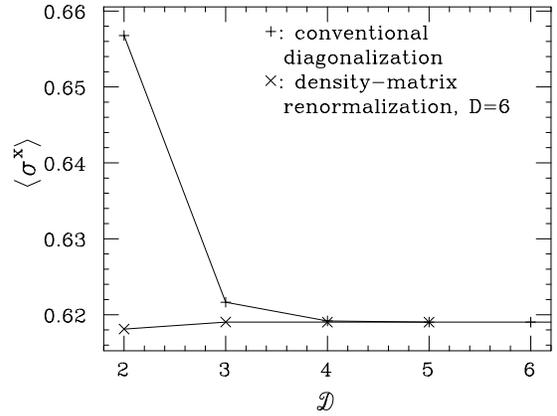}
\end{center}
\caption{
The transverse magnetization is plotted for
the system with $N=8$, $\Delta=0.3$ and $\alpha=0.5$
with the dimensionality ${\cal D}$ of each oscillator varied.
The plots $+$ are evaluated with the conventional
occupation-number representation.
The occupation number is restricted within $n \le n_{\rm max}$.
Hence, ${\cal D}=n_{\rm max}+1$.
The plots $\times$ are those evaluated with
the density-matrix-truncation algorithm.
The dimensionality of each small oscillator $d$ is
varied (${\cal D}=d$) with
the big oscillator dimensionality fixed ($D=6$).
}
\label{shusoku_t5}
\end{figure}

In this subsection, we show the precision of the
algorithm explained in the preceeding subsection.
In Fig. \ref{shusoku_t5},
we plotted the transverse magnetization
$\langle \sigma^x \rangle$ 
for the system
$N=8$, $\Delta=0.3$ and $\alpha=0.5$
by means of the conventional
truncation scheme; namely, the boson states are represented 
in term of the occupation number, and 
the occupation number is 
restricted within $n \le n_{\rm max}$.
(Note that the transverse magnetization indicates
a degree
to what extent the tunneling is disturbed by
the reservoir fluctuations.)
Hence, the dimensionality of each oscillator is given by 
${\cal D}=n_{\rm max}+1$.
We observe that
through increasing ${\cal D}$, the results
converge to a certain limit.
In the same plot, we show the results by means of the
density-matrix-renormalization method 
for the same system as the above ($N=8$, $\Delta=0.3$ and
$\alpha=0.5$).
In this renormalization, we fixed the dimensionality
of the big oscillator as $D=6$, and varied 
the small-oscillator dimensionality $d$;
hence, ${\cal D}=d$.
We see that, with fewer number of bases, the renormalization
results converge more rapidly than the former
occupation-number-%
representation results.
As is mentioned in the previous subsection,
the equilibrium position of each
oscillator is shifted by the coordinate-coordinate coupling.
The relevant bases of the density-matrix renormalization 
are constructed so as to take into account of the fluctuations
from the equilibrium position,
whereas the occupation-number bases are rather inefficient
to represent these shifted oscillator modes.

\begin{figure}[htbp]
\begin{center}\leavevmode
\epsfxsize=8.5cm
\epsfbox{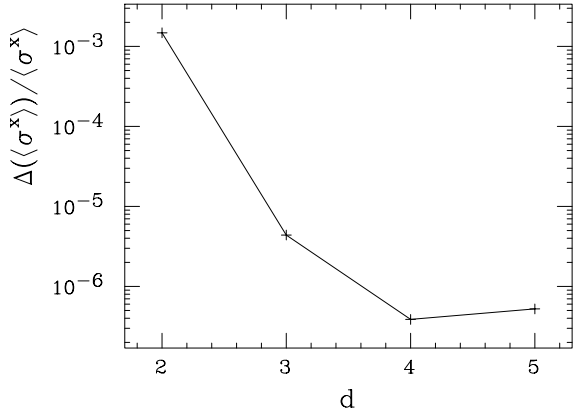}
\end{center}
\caption{
The relative error of the transverse magnetization
with the density-matrix-renormalization
method. We varied the number of remained bases $d$
with $D=6$ fixed.
The system parameters are
the same as those shown in Fig. {\protect \ref{shusoku_t5}}.
}
\label{gosa_t5}
\end{figure}

\begin{figure}[htbp]
\begin{center}\leavevmode
\epsfxsize=8.5cm
\epsfbox{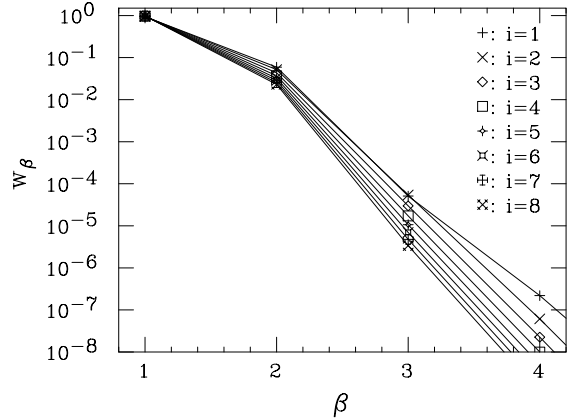}
\end{center}
\caption{
The weight $w^\beta$ (density-matrix eigenvalue) is
plotted for each oscillator mode 
($i=1\sim8$).
}
\label{weight_t5}
\end{figure}

In Fig. \ref{gosa_t5}, we show the relative error of the 
transverse magnetization of the density-matrix-%
renormalization method ($D=6$);
the error is defined as the deviation from the
the conventional occupation-number-truncation 
diagonalization with ${\cal D}=n_{\rm max}+1=6$.
The system parameters are the same as those shown in
Fig. \ref{shusoku_t5}.
We observe, with very limited number of relevant bases ($d=2\sim3$),
the density-matrix renormalization 
reproduces the full-diagonalization result
precisely.
The relative error is saturated to $\sim 10^{-7}$ due to
the numerical round-off error for $d \ge 4$.
The result indicates that a few relevant oscillator modes
are of importance.
In Fig. \ref{weight_t5},
we show the weight $w_\beta$ of each eigenstate ($i=1 \sim D$) for
each reservoir oscillator ($\alpha=1 \sim N$).
The weight $w_\beta$ indicates the statistical weight 
of the state ${\bf u}^\beta$ contributing
to the ground state.
We notice that, in fact, the
first few bases are particularly weighted, and 
the other states are irrelevant ($w_\beta<10^{-5}$),
and to be ignored.
Furthermore, it should be noted that these features are  
common to all the reservoir modes ($i=1 \sim N$).

To summarize, in Fig. \ref{shusoku_t5}, we found that
by means of the conventional occupation-number truncation,
the result converges gradually as the 
occupation-number threshold $n_{\rm max}$ is increased.
We see that the occupation numbers exceeding $n \approx 6$
are hardly excited.
In the density-matrix renormalization, see Figs. 
\ref{gosa_t5}-\ref{weight_t5},
we found that
only the first few states are particularly weighted.
Hence,
hereafter,
we set the dimension of the big site to be $D=6$, and
remain two relevant states ($d=2$) for
each oscillator mode.
We believe that the number of the reservoir oscillators is
prior to
the number of remained bases $d$ for each mode:
Note that as the number of the reservoir modes is increased,
the coupling constants $\{ f_i \}$ should be reduced
correspondingly; the oscillators become less disturbed.
In the following, we simulate the reservoir 
consisting of eighteen oscillator modes.
Hence, the truncation error is expected to be reduced
furthermore from those shown in this subsection.

\section{
Density-matrix-renormalization analysis of
the dissipative tunneling}
\label{section3}

With use of the method developed in the preceeding section,
we investigate the relaxation properties in the ground state
of the
spin boson model (\ref{Hamiltonian}).
These properties are extracted from the dynamical susceptibility,
which is readily evaluated in our scheme.
Our results of the relaxation coefficients
are contrasted with those obtained by means of
the quantum Monte-Carlo simulation \cite{Voelker98}.

\subsection{Dynamic susceptibility --- 
continued-fraction-expansion formula}

In this subsection, we evaluate the dynamical susceptibility,
and compare ours with that obtained at an integrable point
$\alpha=0.5$.
The analyticity of the susceptibility is analyzed extensively
in the next subsection so as to yield relaxation properties.

Linear response theory states that the relaxation
(non-equilibrium) process should be described in term of 
a certain ground-state equilibrium dynamical correlation function,
unless the process is not so far from equilibrium.
In other words, equilibrium dynamical correlation function
does contain informations about non-equilibrium processes. 
For that purpose,
we calculated the following dynamical susceptibility,
\begin{eqnarray}
\chi''(\omega)&=&\frac12
\int_{-\infty}^{\infty}{\rm d}t
{\rm e}^{{\rm i}\omega t}
\langle [\sigma^z(t),\sigma^z]\rangle  \nonumber \\
\label{Czz}
 &=&
{\rm Im}\left(
\left\langle\sigma^z\frac{1}{\omega+E_{\rm g}-{\cal H}+{\rm i}\eta}\sigma^z
\right\rangle   \right. \nonumber \\
\label{dynamical_susceptibility}
& & \ \ \ \ \ 
-
\left.
\left\langle\sigma^z\frac{1}{\omega-E_{\rm g}+{\cal H}-{\rm i}\eta}\sigma^z
\right\rangle
        \right)     \\
&=&{\rm Im}G(\omega) .
\end{eqnarray}
Some might wonder that the inverse matrix of the total Hamiltonian
appearing 
in eq. (\ref{dynamical_susceptibility}) cannot be computed; this is true.
The {\em expectation value} of the inverse 
of the Hamiltonian is, however,
evaluated with use of the Gagliano-Balseiro 
continued-fraction formula \cite{Gagliano87,Gagliano88},
\begin{equation}
\label{Gagliano_Balseiro}
\left\langle f_0 
\left| \frac{1}{z-{\cal H}} 
\right| f_0 \right\rangle
=
\frac{\langle f_0 | f_0 \rangle}
{
z-\alpha_0-\frac{\beta_1^2}
{
z-\alpha_1-\frac{\beta_2^2}
     { z-\alpha_2-\frac{\beta_3^2}{\ddots} }
}
},
\label{Gagliano-Balseiro}
\end{equation}
where the coefficients are given by the
Lanczos tri-diagonal elements,
\begin{eqnarray}
|f_{i+1}\rangle &=& {\cal H}|f_i\rangle-\alpha_i|f_i\rangle
                      -\beta_i^2|f_{i-1}\rangle,           \\
\alpha_i &=& \langle f_i|{\cal H}|f_i\rangle/\langle f_i|f_i\rangle , 
                                                 \nonumber \\
\beta_i^2 &=& \langle f_i|f_i\rangle/\langle f_{i-1}|f_{i-1}\rangle
\ \ \   (\beta_0 = 0). \nonumber
\end{eqnarray}
Therefore,
through choosing of the Lanczos initial vector
as $|f_0\rangle=\sigma^z |\Psi_0\rangle$,
we readily evaluate the dynamical susceptibility
by means of the same numerical procedure
used in the preceeding Lanczos diagonalization.

We expanded the formula (\ref{Gagliano-Balseiro})
up to the one-hundredth order.
This is comparable with the iteration number needed to
obtain the low-lying states in the Lanczos diagonalization.
Therefore, the expansion is expected to yield precise result
concerning low-lying frequencies.
This frequency range is sufficient for our purpose, because
we are concerned in the frequency range $\sim \Delta_{\rm eff}$. 
Moreover,
in practice,
at high frequencies, the magnitude of the
spectral intensity is suppressed considerably.

\begin{figure}[htbp]
\begin{center}\leavevmode
\epsfxsize=8.5cm
\epsfbox{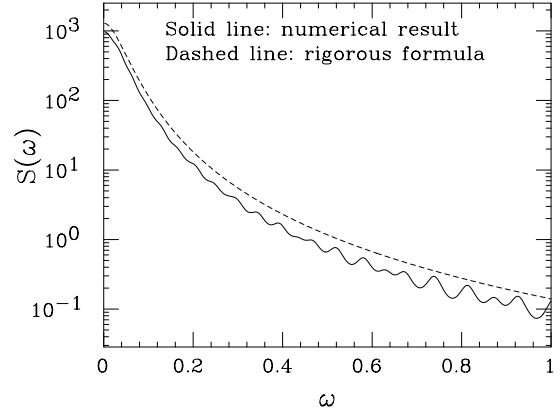}
\end{center}
\caption{
The spectral function $S(\omega)$
({\protect \ref{spectral_function}})
is plotted for $\Delta=0.2$ and $\alpha=0.5$.
The solid line is our density-matrix-renormalization result
for $N=18$. The delta-function peaks are broadened into the
Lorentz form with the width $\eta=0.022$.
The dashed line shows a rigorous result
({\protect \ref{Toulouse}}), which is available at the
Toulouse point $\alpha=0.5$.
}
\label{S_t2_t5}
\end{figure}

In Fig. \ref{S_t2_t5}, we plotted the spectral function,
\begin{equation}
\label{spectral_function}
S(\omega)=\frac{\chi''(\omega)}{\omega}  ,
\end{equation}
for the parameters $\Delta=0.2$, $\alpha=0.5$ and $N=18$.
(The spectral function is related closely to the
linear-response function.)
At the point $\alpha=0.5$, the Hamiltonian (\ref{Hamiltonian})
is reduced to being quadratic through a canonical
transformation, and the spectral intensity
is calculated exactly in the form \cite{Guinea85b},
\begin{eqnarray}
\label{Toulouse}
S(\omega) &=& \frac{8\tilde{\Delta}^2}{\pi}
\frac{1}{\omega^2+4\tilde{\Delta}^2}     \nonumber \\
& &
\times
\left(
\frac{1}{\omega\tilde{\Delta}}\arctan\left(\frac{\omega}{\tilde{\Delta}}\right)
+
\frac{1}{\omega^2}\ln\left( 1+\frac{\omega^2}{\tilde{\Delta}^2} \right)
\right),
\end{eqnarray}
($\tilde\Delta=(\pi/4)\Delta^2/\omega_{\rm c}$).
The rigorous result is plotted in Fig. \ref{S_t2_t5} as well.
We observe that our numerical data reproduces
the rigorous formula.
The slight difference may be attributed to the
spectral form (\ref{ohmic_condition})
of the reservoir around the cut-off frequency.
We used a non-analytic
reservoir spectrum which vanishes suddenly at the
cut-off frequency.
This difference may become irrelevant, if the
tunneling amplitude is set to be sufficiently small compared with
the cut-off frequency.

\subsection{Analyticity of the dynamical susceptibility
(resolvent) and the
damping properties}
\label{section3_2}

\begin{figure}[htbp]
\begin{center}\leavevmode
\epsfxsize=8.5cm
\epsfbox{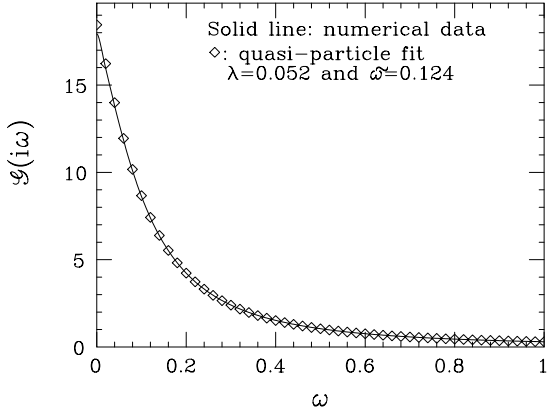}
\end{center}
\caption{
The imaginary-time Green function ({\protect \ref{ondo_G}})
is plotted for $N=18$, $\Delta=0.2$ and $\alpha=0.2$.
The solid line is our density-matrix-renormalization result.
The plots $\diamond$ are those of the `quasi-particle' form
({\protect\ref{quasi_particle}}) with the
damping coefficients $\lambda=0.052$ and $\tilde{\omega}=0.124$.
}
\label{c_t2_t2}
\end{figure}

\begin{figure}[htbp]
\begin{center}\leavevmode
\epsfxsize=8.5cm
\epsfbox{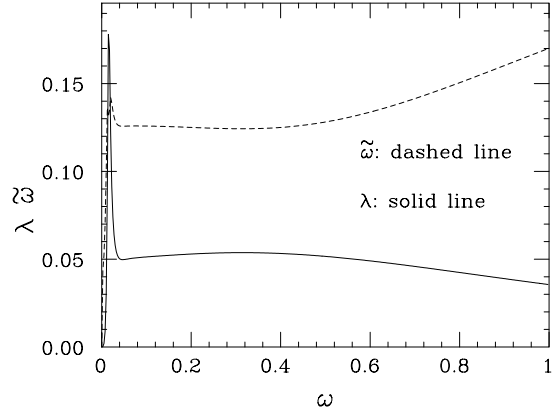}
\end{center}
\caption{
The damping coefficients $\lambda$ ({\protect\ref{lambda_omega}})
and $\tilde{\omega}$ ({\protect\ref{omega_omega}}) are
plotted; the system parameters are the same as those
in Fig. {\protect \ref{c_t2_t2}}.
Around the frequency range $\sim\Delta_{\rm eff}(\approx0.13)$,
these quantities remain invariant, indicating that
over this range of time, the dynamics is
described by the quasi-particle picture
({\protect\ref{quasi_particle}}).
}
\label{damp_t2_t2}
\end{figure}

In this subsection, we investigate the analyticity of the
dynamical susceptibility (\ref{dynamical_susceptibility}),
from which we extract
informations about relaxation properties.
In Fig. \ref{c_t2_t2},
we plotted the imaginary-time correlation function,
\begin{eqnarray}
\label{ondo_G}
{\cal G}({\rm i}\omega) &=& -
  \int_0^\beta{\rm d}\tau \langle 
     {\rm e}^{\tau{\cal H}}\sigma^z{\rm e}^{-\tau{\cal H}}\sigma^z
        \rangle_\beta {\rm e}^{{\rm i}\omega \tau} \\
       &=& G({\rm i}\omega) \ \ \ \ (\beta\to\infty),
\end{eqnarray}
for the system with $\Delta=0.2$, $\alpha=0.2$ and $N=18$.
As is shown in the plot, the numerical result is fitted
well by the `quasi-particle' formula,
\begin{equation}
{\cal G}({\rm i}\omega)=
\frac{A}{(\omega+\lambda)^2+\tilde{\omega}^2}.
\label{quasi_particle}
\end{equation}
This fact was pointed out by V\"olker, who
performed a quantum Monte-Carlo simulation; see Introduction.
As is apparent from the definition (\ref{dynamical_susceptibility}),
the (fitting) parameters $\lambda$ and $\tilde{\omega}$
give the damping rate and the frequency, respectively,
of the `coordinate' $\sigma^z$ perturbed from the equilibrium position.
According to the formula (\ref{quasi_particle}),
the following relations hold;
\begin{eqnarray}
\label{lambda_omega}
\lambda&=&
\frac
{\frac{{\rm d}}{{\rm d}\omega}\left({\cal G}({\rm i}\omega)\right)^{-1}}
{\frac{{\rm d}^2}{{\rm d}\omega^2}\left({\cal G}({\rm i}\omega)\right)^{-1}} 
-\omega  ,  \\
\label{omega_omega}
\tilde{\omega}&=&  
\sqrt{
\frac{2
({\cal G}({\rm i}\omega))^{-1}}
{\frac{{\rm d}^2}{{\rm d}\omega^2}
\left({\cal G}({\rm i}\omega)\right)^{-1}}
-(\omega+\lambda)^2  .
}
\end{eqnarray}
From these relations, we estimate the damping coefficients
$\lambda$ and $\tilde{\omega}$.
It is one of the advantages of our scheme
that one can differentiate the function ${\cal G}({\rm i}\omega)$,
because it is expanded in the analytic form 
(\ref{Gagliano_Balseiro}).
in Fig. \ref{damp_t2_t2},
we plotted the right hand side of eqs. (\ref{lambda_omega}) 
and (\ref{omega_omega});
the system parameters are the same as those in Fig. \ref{c_t2_t2}. 
We see that these quantities are invariant actually over a certain
frequency range $\sim \Delta_{\rm eff}(\approx0.13)$, and thus the 
quasi-particle picture (\ref{quasi_particle}) holds
in the time range $\sim\Delta_{\rm eff}^{-1}$.
As is explained in Introduction, we are concerned in the
physics of this particular range of time.
In consequence, we obtained
the poles of $G(\omega)$ at 
$\omega=\pm \tilde{\omega} - {\rm i}\lambda$, which are {\em away}
from the real axis.
It is noteworthy that
such irreversible features are not transparent in 
the original high-order-expansion result
(\ref{Gagliano_Balseiro}).
Through the above data analysis, such the relaxation
features become clear;
we discuss this point afterwards.

\begin{figure}[htbp]
\begin{center}\leavevmode
\epsfxsize=8.5cm
\epsfbox{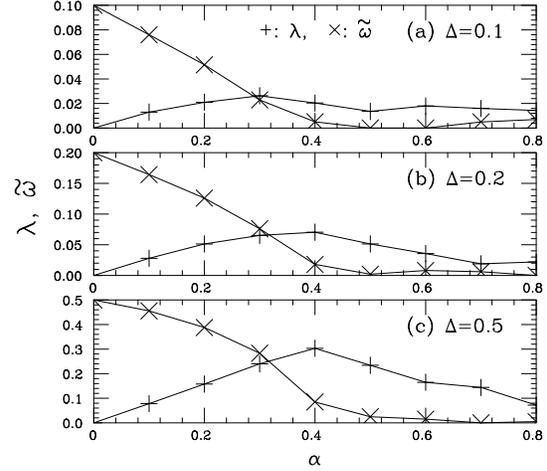}
\end{center}
\caption{
The damping rate $\lambda$ and the
frequency $\tilde{\omega}$ estimated
by means of
the density-matrix-renormalization algorithm;
(a) $\Delta=0.1$, (b) $\Delta=0.2$ and (c) $\Delta=0.5$.
These are to be contrasted with those of
quantum Monte-Carlo method
{\protect \cite{Voelker98}}.
}
\label{kekka}
\end{figure}

In Fig. \ref{kekka}, we plotted the damping rate and the
frequency.
These damping coefficients are estimated both with the
procedure as is shown in Fig. \ref{damp_t2_t2} and with
the conventional least square fit by the
function (\ref{quasi_particle}).
Our results of the density-matrix renormalization
confirm the former
quantum Monte-Carlo study \cite{Voelker98}:
The oscillation frequency $\tilde{\omega}$
is suppressed as the dissipation is strengthened.
It vanishes around $\alpha\approx0.5$.
This point indicates the transition 
between the under-damped and the over-damped oscillations.
By means of the bosonization technique \cite{Guinea85},
this transition point was predicted to locate at $\alpha=0.5$
irrespective of the tunneling amplitude strength $\Delta$.
Because in the bosonization technique, the band width (cut-off
frequency) is supposed to be sufficiently large
compared with the many-body interactions ($\Delta$ and 
$\alpha$),
the validity in the  strong-coupling region
is not necessarily assured.
In Fig. \ref{kekka}, in fact,
we see that the $\tilde{\omega}$ plots become curved convexly,
as the tunneling amplitude is strengthened, and accordingly
it becomes evident that the transition
point locates below $\alpha=0.5$.
These features were reported in the paper \cite{Voelker98},
and were suspected to be due to a systematic numerical error
caused by the critical slowing down.
In our diagonalization calculation, we are free from 
the critical slowing down.
Therefore, we conclude that for larger values of $\Delta$,
the transition point actually shifts below $\alpha=0.5$.
Furthermore, we confirm the report \cite{Voelker98}
that at $\alpha=1/3$, the damping rate and the
frequency coincide.
At this point, the peaks around $\omega=\pm\tilde{\omega}$ of
spectral intensity (\ref{spectral_function})
merge into a single peak
so that this point was suspected 
to indicate a certain 
phase transition.
The present result suggests that the point is merely the situation 
where
the damping feature smears out the oscillatory behaviour,
because these relative strengths exchange.

We see that in Fig. \ref{kekka} (a) ($\Delta=0.1$),
the plots for $\alpha>0.6$ are rather scattered (irregular).
In that region, the quasi-particle poles shift
towards the origin of the complex plane 
($\pm\tilde{\omega}-\lambda\to0$),
so that the dumping parameters degenerate into the
slowest reservoir oscillation mode.
In that situation, our method becomes inapplicable.
Similar difficulty arises in the vicinity of the
localization transition point $\alpha=1$ for larger values
of $\Delta$.

We are in a position to discuss the above quasi-particle feature
furthermore in detail.
It is noteworthy
that the form (\ref{quasi_particle}) implies that the
analyticity of the dynamical susceptibility
is broken along
the real axis; the upper and lower complex planes are
not continued analytically.
This is precisely due to the reservoir effect, which drives the spin state
to be in equilibrium, violating the
time-reversal symmetry.
In our numerical result, however,
the upper and the lower complex planes are continued analytically,
although
along the real axis, 
there distribute vast number of poles densely.
This feature contradicts the above quasi-particle picture
insisting isolated poles at $\omega=\pm\tilde\omega-{\rm i}\lambda$.
This inconsistency is simply due to the fact that
our reservoir spectrum is not continuous, and thus
the time-reversal symmetry is maintained.
Only through the limit of infinite oscillator modes,
these poles merge into the `quasi-particle' poles away from
the real axis.
Hence, in order to extract relaxation properties,
one should avoid the real axis, along which 
the result is suffered significantly
from discontinuity of the reservoir modes.
Along the imaginary axis, as is shown in Fig. \ref{c_t2_t2},
the result is smooth and monotonic,
and is much easier to be fitted
by the quasi-particle form.
That is why we examined the imaginary-time Green function (\ref{ondo_G})
just as V\"olker did for analyzing his Monte-Carlo data
computed at each Matsubara frequency.
We stress that
our resolvent
is evaluated (expanded) in the analytical 
(continued-fraction) form as in eq. (\ref{Gagliano_Balseiro}).
Therefore, as is shown previously in Fig. \ref{S_t2_t5},
one can obtain
the spectral function (real-time Green function) immediately
through performing the Wick rotation.
This is one of the advantages of the present scheme
over others.

\section{Summary and discussions}
\label{section4}

We investigated the dissipative two-state system 
(\ref{Hamiltonian}) through applying
the density-matrix Hilbert-space-truncation algorithm 
\cite{Zhang98}.
An implementation of this algorithm
is 
proposed,
and
the precision applied to the present model
is reported in detail.
We found that in our situation where the oscillator equilibrium
position is shifted by the coordinate-coordinate coupling,
the density-matrix renormalization works much better than 
the conventional occupation-number-truncation method.
We have remained two relevant states for each oscillator mode,
so that we succeeded in treating eighteen oscillators, keeping
the truncation error considerably small.
We belive that the number of tractable oscillators is more crucial
in observing the relaxation properties than the
`quality' (fidelity) of
each oscillator mode.
In fact, we reproduced the dynamical susceptibility
at the Toulouse point $\alpha=0.5$, at which
a rigorous formula is known.
We stress that, in our scheme, the susceptibility is evaluated
in the {\em analytical} continued-fraction form (\ref{Gagliano_Balseiro}),
which is apparently advantageous over others in
performing the analytic continuation (Wick rotation);
in order to observe relaxation (time irreversible) properties
from numerical data, which apparently possesses the
time-reversal symmetry,
we need to examine the analyticity of the resolvent 
(dynamical susceptibility); see Section \ref{section3_2}. 
In fact,
after the immediate
analytic continuation to the imaginary axis,
we found that the result is well fitted by the form (\ref{quasi_particle}),
confirming V\"olker's finding \cite{Voelker98}
with quantum Monte-Carlo method.
That is, the susceptibility possesses the so-called
quasi-particle poles away from the real axis 
$\omega=\pm\tilde{\omega}-{\rm i}\lambda$.
In consequence, we obtained
the damping rate $\lambda$ and the frequency $\tilde{\omega}$.
Both
agree with those of the
quantum Monte-Carlo method \cite{Voelker98}.
In particular,
our results confirm the former report that for larger values of
$\Delta$, the transition point between the over-damped and
the under-damped oscillations shifts downwards ($\alpha_{\rm c}<0.5$).
For studying such critical property, our method
is advantageous over the Monte-Carlo method,
because our method is free from the critical slowing down.

\section*{Acknowledgement}
Our program is based on the subroutine package TITPACK ver. 2
coded by professor H. Nishimori.

\end{document}